\documentstyle[11pt,aaspp4]{article}
\slugcomment{To Appear in August 1997 PASP}

\lefthead{Gizis \& Reid}
\righthead{LHS Stars}

\begin{document}

\title{Probing the LHS Catalog. I. New Nearby Stars and the Coolest Subdwarf
\footnote
{Observations were made partially at the 60-inch telescope at Palomar
Mountain which is jointly owned by the California Institute of Technology
and the Carnegie Institution of Washington}
 }

\author{John E. Gizis}
\author{I. Neill Reid\altaffilmark{2}}
\affil{Palomar Observatory, 105-24, California Institute of Technology,
  Pasadena, California 91125, e-mail: jeg@astro.caltech.edu, 
  inr@astro.caltech.edu}

% affiliations, which are identified by the \altaffilmark after each name. 

\altaffiltext{2}{Visiting Research Associate, Carnegie Institute of Washington}

\begin{abstract}

We present moderate resolution spectroscopy of 112 cool dwarf stars to
supplement the observations we have already presented in 
the Palomar/MSU Nearby-Star Spectroscopic Survey.  
The sample consists of 72 suspected nearby stars added to the
{\it The Preliminary Third Catalog of Nearby Stars} since 1991 as well
as 40 faint red stars selected from the LHS catalog.
LHS 1826 is more metal-poor and cooler than the coolest previously known
extreme subdwarf, LHS 1742a.  LHS 2195 is a very late 
M dwarf of type ${\rm M} 8~{\rm V}$, probably at a distance
of ten parsecs.  LHS 1937 is an ${\rm M} 7~{\rm V}$ star at
$\sim 20$ parsecs.  Three other previously unobserved
LHS stars have estimated distances that place them within 25 parsecs.  

\end{abstract}

%\keywords{}

\section{Introduction \label{intro}}

M dwarfs make up the vast majority ($\sim 80\%$ by number) of the 
stellar component of the Galaxy.  With the exception of the
few clusters which are near enough to study, field samples 
are our primary source of observational data on these cool stars.  
The two most important compilations are the LHS Catalog (\cite{l79}),
listing the known stars with $\mu > 0.5$ arcseconds per year,
and the preliminary Third Catalog of Nearby Stars (\cite{gj91}, 
hereafter pCNS3),
listing the known stars within 25 parsecs.  There is considerable
overlap between these two samples, and indeed most of the nearby stars were
first identified in proper motion surveys.  Despite the importance of
the proper motion stars, there have been no followup observations of
many of the fainter LHS stars. 

Spectroscopy is a powerful way to make an initial survey of 
these stars because it
measures temperature, metallicity, and chromospheric activity.
We have already presented the first stage of the Palomar/Michigan
State University  
Nearby Star Spectroscopic Survey (\cite{rhg95}, hereafter PMSU1;
\cite{hgr96}, hereafter PMSU2), which includes measurements of 
the 2063 accessible pCNS3 M dwarfs over the whole
sky in a consistent system.  These observations were used to
identify misclassified and metal-poor stars (PMSU1, PMSU2),
estimate the completeness of the catalog (PMSU1),
redetermine the disk luminosity function (PMSU1), 
reanalyze the local disk space kinematics (PMSU1),
and investigate the dependence of chromospheric activity on spectral type
(PMSU2).  In addition, Gizis (1997, hereafter G97) has obtained 
spectra on the same system for a large sample of cool 
metal-poor LHS stars.  These observations 
were used to develop a spectral classification system and 
determine the metallicity scale for the M subdwarfs.  

Here we present spectroscopic observations on this same system 
aimed at identifying interesting red dwarfs, in particular
new nearby stars, ultra-cool dwarfs, and very metal-poor stars.  
In Section~\ref{data}, we review the sample selection and data reduction.  
In Section~\ref{nearby}, we discuss observations of a sample of suspected
nearby stars added to pCNS3 since 1991.  In Section~\ref{faintlhs},
we present observations of selected faint ($r > 17$) LHS 
stars that have been previously neglected.  We summarize
in Section~\ref{conclusions}.

\section{Sample Selection and Data Reduction \label{data}}

Two samples of stars were selected for observations.  First,   
Dr. Jahrei{\ss} provided us with a list of 85 candidate
nearby stars (53 of which are LHS stars) 
which were not included in pCNS3.  Most are relatively
bright ($V < 15$) and have been identified by 
(B)VRI photometry; a handful have trigonometric parallaxes of
low quality.  We have
observed 72 of these stars as well as one common proper motion companion
(G240-045) not included in the list.  Second, we have
begun a program of observing faint ($m_r > 17$) red
LHS stars without published data.  Dr. Kirkpatrick
kindly provided a list of previous unpublished observations to prevent
duplication.   A few of our targets do have some previously
published observations; in particular, we note that  
LHS 1506, 3868, 5356, 5359, and 5360 have accurate USNO 
trigonometric parallaxes (\cite{m92}).  
After taking the observations, we found that 
spectra of LHS 1826, 2099, and LHS 2100 were presented by
Bessell \& Scholz (1990). 

Our observations were taken and reduced in the same way as described in
PMSU1, PMSU2, and G97.  The new nearby star candidates
were observed at the Palomar 60-in. in June 1995 and January 1996,
at the Hale 200-in. telescope in October 1995, and at the Las Campanas  
100-in. telescope in December 1995.  The faint red LHS stars were
observed at the Hale 200-in. in November 1996 and the Las Campanas 100-in.
in December 1996.  The resolution was $\sim 3 \AA$ per pixel with
wavelength coverage from $6200 - 7500\AA$.  Marcy \& Benitz (1989) 
stars were used as radial velocity standards.  No radial velocities
were determined for the December 1996 data.  Radial velocities 
have an accuracy of $\sim 20 {\rm ~km~s}^{-1}$.  The bandstrengths 
are accurate to $\sim 0.02$ with the resulting spectral types 
accurate to 0.5 subclasses.   

\section{The Candidate Nearby Stars\label{nearby}}

We present the data for the nearby dwarfs in Tables~\ref{nearby-data},
~\ref{nearby-bands}, and ~\ref{nearby-uvw} which are in the same format as 
Tables 1, 2, and 3 of PMSU1 and PMSU2.  
Positions in Table~\ref{nearby-data} were redetermined
for the stars using the Digitized Sky Surveys with proper
motions added to make the positions epoch J2000.
The spectra show that the candidate nearby stars are all ordinary disk
dwarfs: none are M subdwarfs and the frequency and strength of
$H \alpha$ emission is consistent with that found in PMSU2.  

The system made up of G 035-027 and LP 469-118 has a photoelectric
V measurement by Fleming et al. (1988).  They were unable to resolve 
the binary and therefore reported V magnitudes by
on the basis of an assumed $\Delta V$ of 2.53 magnitudes.
That value was derived from their own spectral types; however, we
predict $\Delta V \sim 0.4$  based on the TiO indices.  
The noted appearance of the system in the TV guider and on the Digitized
Sky Survey suggests only a small magnitude difference, and 
in fact, Luyten estimated $\Delta m_r \sim 0.1$.  We would like to note
however that our observations were taken a large airmass and that 
the G 035-027 spectrum has a very red slope, which we attribute to 
an observational artifact.   We estimate the distance in 
Tables~\ref{nearby-data} and~\ref{nearby-uvw} assuming the
total magnitude ($V=14.20$) reported by Fleming {\it et al.} 
and our own $\Delta V$ of 0.4.  

None of the candidate nearby stars appear to belong in the eight parsec sample,
but one may lie within ten parsecs.  LHS 2395
is 9.9 parsecs away on the basis of the 
TiO index.  The same value is derived on the basis of its color
in the Jahrei{\ss} list.  

\section{The Faint LHS Stars\label{faintlhs}}

Indices, radial velocities, and spectral types (G97 system) for the faint
LHS stars are given in Table~\ref{lhs-spec}.  In this 
sample, 19 are near-solar metallicity M dwarfs (M~V),
13 are sdM, and 6 are esdM.  We also attempted to observe
LHS 5082 but were unable to detect it with either the 
Double Spectrograph on the 200-in. or VRI CCD imaging using
the Palomar 60-in. telescope.  On the Palomar Sky Survey images, 
a diffraction spike from 
the much brighter LHS 5081 passes through the position listed 
by Luyten.  We conclude that LHS 5082 does not exist.  

\begin{figure}
\plotone{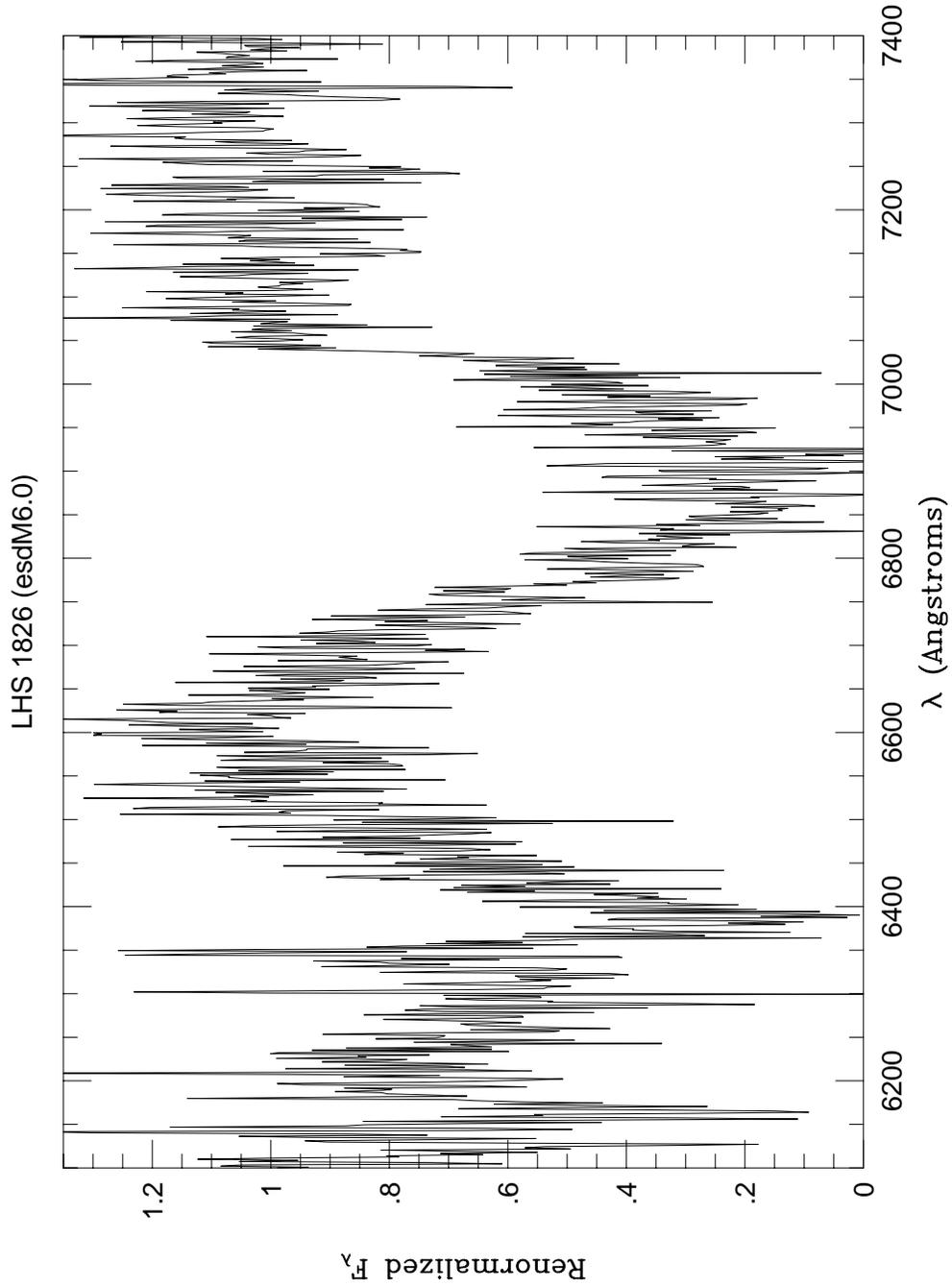}
\figcaption{Spectrum of LHS 1826. The CaH indices give a spectral type of
esdM6.0 on the G97 system and the lack of TiO implies this star is more 
metal poor than most esdM.    
\label{lhs1826}}
\end{figure}

LHS 1826 is the coolest extreme subdwarf known.  A spectrum appears
in Bessell \& Scholz (1990), but our indices allow us to estimate 
a quantitative spectral type of esdM6.0, later than either
the parallax stars LHS 3061 (esdM5.0, $M_V = 14.25$) or
LHS 1742a (esdM5.5, $M_V = 14.44$).  The spectrum is plotted in
Figure~\ref{lhs1826}.  
The value for TiO5 should be viewed with caution because the
spectrum is noisy, but
there is no doubt that the CaH bands are stronger and the TiO weaker
than either LHS 3061 or 1742a.  It appears LHS 1826, like LHS 453, is
more metal-poor than most extreme subdwarfs.  Fits to the Allard
\& Hauschildt (1995) synthetic spectra using the G97 procedure
gives $T_{eff} = 3300$ and $[m/H] = -2.5$, although fits with
cooler temperatures and lower metallicities are almost as good.
There is evidence of atomic lines in the spectrum which supports
$[m/H] \sim -2.5$, but there is too much noise to measure them.    

LHS 2099 and 2100 are a pair of extreme subdwarfs with common proper motion.  
LHS 2100 appears to be signifcantly fainter than expected given the 
similarity in the CaH strength (i.e., temperature)  when compared
with LHS 2099.  No accurate photometry is
available, but the LHS catalog lists 
$\Delta m_r = 3$ and comparison of the flux in our spectra gives
$\Delta m_r \sim 2$.  A linear fit to the extreme subdwarf sequence
in the HR diagram suggests that the luminosity difference between the
two stars should only be $\sim 0.2$ magnitudes.  If LHS 2099 is an
unresolved equal-luminosity binary, then we would expect an
additional 0.75 magnitudes difference, but this is not enough to
explain the observations.  The system
may be similar to GJ 1230 which consists of stars that lie
at the top and bottom of the ``kink'' in the HR diagram at spectral
type M4.0~V (\cite{gr96}).  At the kink, a small difference
in spectral type or color corresponds to a large difference in
luminosity; this behavior is not accounted for in a simple linear fit.   
For metal-poor stars, there may be a similar kink 
in the HR diagram at spectral type $\sim esdM2.0 - 2.5$.
Comparing to the colors and spectral types of 
the extreme subdwarfs in G97, we find that 
LHS 2099 and 2100 should have $V-I \sim 2.0$.
Accurate photometry is needed to measure the luminosity 
ratio and colors of these two stars.  

Another subdwarf system of interest is LHS 2139/2140.  For this
observation, the spectrograph slit was rotated to the position of 
angle of $24\deg$ given by Luyten.  The primary, LHS 2140, is
an sdM0.5 but the fainter LHS 2139 is not an M-star.  
Our spectrum is noisy but appears to be featureless; the ratio of fluxes
in the spectrum gives $\Delta m_r \sim 4.7$ which is consistent with
Luyten's value $\Delta m_r \sim 4.6$.   We conclude
that LHS 2139 is a cool white dwarf.  LHS 2140 is spectroscopically
very similar to LHS 307 (G97) which implies LHS 2140 is at $\sim 80$
parsecs:  the implied $M_r \sim 14$ for LHS 2139 is reasonable for an
old halo white dwarf.  

Until now, very few cool sdM have been identified (\cite{m92}; G97).
However, Table~\ref{lhs-spec} 
lists eight stars with spectral types in the range sdM4.0
to sdM6.5. The previously observed lack of these stars has thus been 
due to chance or selection effects.  Photometry and trigonometric parallaxes  
would fill in the present gap in the HR diagram.  

LHS 2195 is an ultra-cool M dwarf.  We have shown in PMSU1 
that our indices are not useful beyond $\sim \rm{M}6.5$, because TiO
and CaH begin to weaken with decreasing temperature.  We
therefore have compared our spectrum by eye to PMSU1 spectra of
very late M dwarfs with spectral types by Kirkpatrick, Henry,
\& McCarthy (1991).  We find a spectral type of M8~V;
observations that include redder wavelengths will be needed 
to obtain a definitive spectral type.   Using the Luyten $m_r$ magnitudes and
comparing to the M8~V star vB 10 (LHS 474), we estimate that the
distance to LHS 2195 is 10 parsecs.  The resulting tangential velocity is
$\sim 43 {\rm ~km~s}^{-1}$, which together with the radial 
velocity of $\sim 50 {\rm ~km~s}^{-1}$,
suggests that LHS 2195 is an old disk star and not a young brown dwarf.  

LHS 1937 is also an ultra-cool M dwarf; in this case, our
Las Campanas spectrum extends to $8000 \AA$ and shows VO absorption.
We estimate that it is spectral type M7.0~V.  Using the Luyten 
r magnitudes and the comparison star vB 8 (LHS 429), 
we estimate a distance of $\sim 21$ parsecs, which implies
$V_{tan} \sim 55 {\rm ~km~s}^{-1}$.  

For the remaining M~V (near-solar metallicity) stars we have estimated
absolute magnitudes and distances using the relation 
$M_r = 19.4 - 16.0 \times \rm{TiO5}$, which is a ``by-eye'' fit to the LHS
r magnitudes for the eight parsec single stars (\cite{gr96}).
As the LHS magnitudes are not very accurate, these estimates are
necessarily quite crude.
Three stars have estimated distances within 25 parsecs:
LHS 1146 (24 pc), LHS 1531 (17 pc), and LHS 2064 (23 pc).  
Another four stars (LHS 1157, 1506, 3762, and 5356) 
have estimated distances between 30 and 35 parsecs, although 
the USNO parallaxes show that LHS 1506 and 5356 actually lie 
at $41.5 \pm 2.2$ and $56.8 \pm 14.2$ parsecs respectively.
Three stars seem to have improbably large distances
(their apparent magnitudes are too faint for their spectral type
given their proper motion):  LHS 1294,
LHS 1657, and LHS 1986.  These stars have most likely been misidentified
by us, or may have been mismeasured by Luyten.  We note that
LHS 1986 is reported as a DC white dwarf in McCook \& Sion (1987).  

\section{Conclusions \label{conclusions}}

Our spectroscopic observations of the candidate nearby stars 
have not found any unusual stars, but do support the photometric estimate that
LHS 2395 lies within ten parsecs.  

Obtaining trigonometric parallaxes or other followup observations 
is demanding on observing time
and often must be restricted to the most promising candidates,  
We wish to bring attention to the faint LHS stars 
that deserve additional followup.  We have identified five stars 
likely to be within 25 parsecs 
(including an M8~V  star), eight M subdwarfs later
than sdM4.0, a possible old halo white dwarf, 
and six extreme M subdwarfs (including an esdM6.0 which
is more metal-poor than the previously known very cool extreme M subdwarfs).  

\acknowledgments

We wish to thank the staffs of Palomar and Las Campanas Observatories.
We also would like to thank Dr. Jarei\ss~ for providing 
the list of candidate nearby stars and making useful comments.
JEG gratefully acknowledges partial support from both a Greenstein 
and a Kingsley Fellowship.   
This research has made use of the Simbad database, operated at
CDS, Strasbourg, France.  
The Digitized Sky Surveys were produced at the Space Telescope 
Science Institute
under U.S. Government grant NAG W-2166. The images of these surveys are
based on photographic data obtained using the Oschin Schmidt Telescope on
Palomar Mountain and the UK Schmidt Telescope. 

%Tables 

\begin{deluxetable}{crrrrrrrcrrrrrlr}
\tablewidth{0pc}
\tablenum{1}
\label{nearby-data}
\tablecaption{Candidate Nearby Stars: Basic Data}
\tablehead{
\colhead{name}      &
\colhead{b}          & 
\colhead{}          & 
\colhead{$\alpha$}  &
\multicolumn{2}{c}{(J2000)} & 
%\colhead{}    &
\colhead{$\delta$}  & 
\colhead{}  &
\colhead{Src.\tablenotemark{a}} & 
\colhead{M$_V$} &
\colhead{r}  & 
\colhead{$\epsilon_r$}  &
\colhead{\%$_\pi$}  & 
\colhead{\%$_S$}  &
\colhead{Sp} &
\colhead{V}
}
\tablecolumns{16}
\startdata
  LHS 1037 &  &    0 &  11 &  56.3 &  33 &   3 &  17  & 5 & 11.72 & 18.0 & 5 & 0 & 100 & M3.5 & 13.00 \nl 
  LHS 1062 &  &    0 &  21 &  53.9 &  38 &  16 &  29  & 5 & 11.56 & 18.1 & 5 & 0 & 100 & M3.0 & 12.85 \nl 
  LHS 1104 &  p  &    0 &  35 &  53.3 &  52 &  41 &  14  & 5 & 10.80 & 22.3 & 5 & 48 & 52 & M2.5 & 12.54 \nl 
  LHS 1105 &  s  &    0 &  35 &  53.7 &  52 &  41 &  39  & 5 & 13.14 & 24.5 & 6 & 52 & 48 & M4.0 & 15.09 \nl 
  G 217-060 &  &    0 &  39 &  18.7 &  55 &   8 &  13  & 5 & 12.73 & 19.4 & 3 & 75 & 25 & M3.5 & 14.17 \nl 
  LHS 1169 &  &    0 &  57 &   2.9 &  45 &   5 &   9  & 5 & 11.26 & 13.7 & 4 & 0 & 100 & M3.0 & 11.95 \nl 
  LHS 1182 &  &    1 &   3 &  14.7 &  71 &  13 &  14  & 5 & 12.08 & 22.7 & 7 & 0 & 100 & M3.5 & 13.86 \nl 
  G 173-018 &  &    1 &  45 &  18.4 &  46 &  32 &  10  & 5 & 10.31 & 16.8 & 5 & 0 & 100 & M2.0 & 11.44 \nl 
  LHS 1311 &  &    1 &  53 &  50.5 & -10 &  32 &  14  & 5 & 14.10 & 18.4 & 6 & 0 & 100 & M5.0 & 15.43 \nl 
  G 035-027 & p  &    2 &   8 &  12.2 &  15 &   8 &  45  & 5 & 12.95 & 59.8 & 18 & 0 & 100 & M4.0 & 14.76\tablenotemark{b} \nl 
  LP 469-118 & s  &    2 &   8 &  12.2 &  15 &   8 &  45  & 5 & 13.37 & 15.4 & 5 & 0 & 100 & M4.5 & 15.18\tablenotemark{b} \nl 
  LP 709-43   &  &    2 &  10 &   3.6 &  -8 &  53 &   0  & 5 & 11.29 & 13.9 & 0 & 91 & 9 & M3.5 & 12.\phantom{0}\phantom{0} \nl 
  LHS 1349 &  &    2 &  10 &  35.5 &  46 &  42 &   5  & 5 & 14.37 & 18.6 & 6 & 0 & 100 & M5.0 & 15.71 \nl 
  LHS 1412 &  &    2 &  30 &  35.0 & -15 &  43 &  25  & 5 & 11.16 & 24.4 & 7 & 0 & 100 & M3.0 & 13.1\phantom{0} \nl 
  G 134-035 &  &    2 &  30 &  44.3 &  40 &  28 &  53  & 5 & 12.12 & 16.4 & 2 & 90 & 10 & M1.0 & 13.19 \nl 
  LHS 1553 &  &    3 &  30 &  55.0 &  70 &  41 &  14  & 5 & 11.56 & 24.9 & 7 & 0 & 100 & M3.0 & 13.54 \nl 
  G 038-024 &  &    4 &   4 &   6.0 &  30 &  42 &  46  & 5 & 10.71 & 25.6 & 4 & 75 & 25 & M1.5 & 12.75 \nl 
  LHS 1631 &  &    4 &   8 &  11.3 &  74 &  23 &   1  & 5 & 11.69 & 21.4 & 6 & 0 & 100 & M3.5 & 13.34 \nl 
  LHS 1659 &  &    4 &  20 &  57.5 &  37 &  28 &  45  & 5 & 10.97 & 22.3 & 5 & 26 & 74 & M3.0 & 12.71 \nl 
  LHS 1675 &  &    4 &  31 &   7.3 &  -5 &  20 &  25  & 5 & 12.96 & 13.0 & 1 & 82 & 18 & M3.5 & 13.54 \nl 
  LHS 1679 &  &    4 &  33 &  17.7 &  68 &  46 &  52  & 5 & 14.21 & 21.9 & 7 & 0 & 100 & M5.0 & 15.91 \nl 
  LHS 1808 &  &    6 &   2 &  25.6 &  66 &  20 &  40  & 5 & 13.24 & 18.1 & 5 & 0 & 100 & M4.5 & 14.52 \nl 
  LHS 1817 &  &    6 &   5 &  29.6 &  60 &  49 &  23  & 5 & 13.04 & 13.5 & 4 & 0 & 100 & M4.5 & 13.69 \nl 
  LTT 2494 &  &    6 &  14 &  57.0 &  -5 &   2 &  53  & 5 & 11.69 & 18.3 & 4 & 54 & 46 & M3.0 & 13.\phantom{0}\phantom{0} \nl 
  LTT 2495 &  &    6 &  15 &   0.5 &  -5 &   2 &  45  & 5 & 10.74 & 22.5 & 3 & 79 & 21 & M0.5 & 12.5\phantom{0} \nl 
  LHS 1853 &  &    6 &  32 &  30.9 &  64 &   6 &  18  & 5 & 12.76 & 18.2 & 5 & 0 & 100 & M4.0 & 14.06 \nl 
  LHS 1962 &  p  &    7 &  58 &  18.5 &  87 &  57 &  40  & 5 & 12.29 & 22.7 & 7 & 0 & 100 & M4.0 & 14.07 \nl 
  LHS 1964 &  &    7 &  58 &  22.1 &  49 &  39 &  54  & 5 & 12.84 & 11.6 & 3 & 0 & 100 & M4.0 & 13.16 \nl 
  LHS 1973 &  &    8 &   2 &  44.0 &  59 &   1 &  43  & 5 & 11.65 & 25.7 & 8 & 0 & 100 & M3.5 & 13.70 \nl 
  LHS 1993 &  &    8 &  13 &  44.7 &  79 &  18 &  12  & 5 & 13.41 & 19.0 & 6 & 0 & 100 & M4.5 & 14.80 \nl 
  LHS 2002 &  &    8 &  21 &  57.1 &  17 &  48 &  56  & 5 & 14.30 & 19.1 & 6 & 0 & 100 & M5.0 & 15.71 \nl 
  LHS 2025 &  &    8 &  31 &  29.8 &  73 &   3 &  47  & 5 & 12.42 & 12.6 & 1 & 79 & 21 & M3.5 & 12.92 \nl 
  LTT 12375 &  &    9 &   3 &  53.3 &  12 &  59 &  27  & 5 & 10.35 & 25.4 & 8 & 0 & 100 & M2.0 & 12.37 \nl 
  G 253-034 &  &   10 &   1 &  11.5 &  81 &   9 &  21  & 5 & 13.48 & 22.3 & 7 & 0 & 100 & M4.5 & 15.23 \nl 
  G 042-036 &  &   10 &   6 &  57.9 &  12 &  40 &  54  & 5 & 9.90 & 26.9 & 8 & 0 & 100 & M1.5 & 12.05 \nl 
  LHS 2232 &  &   10 &  12 &  34.5 &  57 &   3 &  50  & 5 & 11.57 & 10.8 & 3 & 0 & 100 & M3.0 & 11.73 \nl 
  Ross 895 &  &   10 &  47 &  24.3 &   2 &  35 &  35  & 5 & 9.77 & 31.0 & 7 & 31 & 69 & M2.0 & 12.23 \nl 
  LHS 2395 &  &   11 &  19 &  30.6 &  46 &  41 &  43  & 5 & 15.81 & 9.9 & 3 & 0 & 100 & M5.5 & 15.78 \nl 
  LHS 2403 &  &   11 &  25 &   0.5 &  43 &  19 &  37  & 5 & 13.37 & 22.0 & 7 & 0 & 100 & M4.5 & 15.08 \nl 
  Wo 9364 &  &   11 &  31 &  36.2 &  40 &  30 &   0  & 5 & 7.81 & 25.0 & 3 & -1 & -1 & K & 9.80 \nl 
  LP 63-218 &  &   11 &  33 &  55.6 &  62 &  22 &   5  & 5 & 12.98 & 29.1 & 9 & 0 & 100 & M4.0 & 15.30 \nl 
  LHS 2439 &  &   11 &  40 &  29.5 &  77 &   4 &  20  & 5 & 10.29 & 19.3 & 6 & 0 & 100 & M2.0 & 11.72 \nl 
  LHS 2513 &  &   12 &   7 &  57.6 &  77 &  15 &   0  & 5 & 11.79 & 21.9 & 7 & 0 & 100 & M3.5 & 13.49 \nl 
  Steph 2.809\tablenotemark{c} &  &   12 &  15 &  38.3 &  52 &  39 &  11  & 5 & 11.52 & 16.1 & 5 & 0 & 100 & M3.0 & 12.55 \nl 
  LHS 2546 &  &   12 &  21 &  29.3 &  54 &   8 &   5  & 5 & 12.31 & 32.0 & 10 & 0 & 100 & M4.0 & 14.83 \nl 
  LP 321-300 &  &   12 &  30 &  55.4 &  31 &  52 &  12  & 5 & 12.18 & 25.8 & 8 & 0 & 100 & M3.5 & 14.24 \nl 
  LHS 2613 &  &   12 &  42 &  49.8 &  41 &  53 &  46  & 5 & 12.15 & 10.8 & 1 & 79 & 21 & M3.5 & 12.31 \nl 
  EXO1259+1238 &  &   13 & 2 & 5.9 &  12 & 22 & 21 & 5 & 10.68 & 26.1 & 8 & 0 & 100 & M2.5 & 12.76 \nl 
  LHS 2672 &  &   13 &   2 &  47.4 &  41 &  31 &   9  & 5 & 11.72 & 17.6 & 5 & 0 & 100 & M3.5 & 12.95 \nl 
  LHS 2686 &  &   13 &  10 &  12.8 &  47 &  45 &  21  & 5 & 13.79 & 14.0 & 4 & 0 & 100 & M4.5 & 14.52 \nl 
  G 014-046 &  &   13 &  20 &  25.0 &  -1 &  39 &  28  & 5 & 9.88 & 22.2 & 4 & -1 & -1 & K & 11.61 \nl 
  LP 271-35 &  &   14 &  31 &  43.1 &  32 &  14 &  32  & 5 & 12.77 & 24.7 & 7 & 0 & 100 & M4.0 & 14.74 \nl 
  LHS 2935 &  &   14 &  32 &   8.5 &   8 &  11 &  31  & 5 & 14.39 & 18.2 & 5 & 0 & 100 & M5.0 & 15.69 \nl 
  LHS 3009 &  &   15 &   0 &  10.2 &  60 &  21 &  47  & 5 & 12.00 & 21.1 & 6 & 0 & 100 & M3.5 & 13.62 \nl 
  LHS 3031 &  &   15 &  10 &  12.0 &  46 &  24 &   7  & 5 & 13.07 & 26.4 & 8 & 0 & 100 & M4.5 & 15.18 \nl 
  LHS 3044 &  &   15 &  14 &  46.8 &  64 &  33 &  43  & 5 & 11.94 & 25.7 & 8 & 0 & 100 & M3.5 & 13.99 \nl 
  GJ 2120 &  &   16 &  27 &  33.2 & -10 &   0 &  29  & 5 & 9.64 & 17.9 & 3 & -1 & -1 & K & 10.90 \nl 
  G 240-044 &  p  &   17 &   6 &  17.7 &  64 &  38 &   8  & 5 & 11.06 & 26.5 & 8 & 0 & 100 & M2.5 & 13.18 \nl 
  G 240-045 &  s  &   17 &   7 &  28.2 &  64 &  37 &  27  & 5 & 12.56 & 29.4 & 9 & 0 & 100 & M4.0 & 14.90 \nl 
  LHS 3318 &  &   17 &  42 &  42.5 &  75 &  37 &  23  & 5 & 13.43 & 15.0 & 4 & 0 & 100 & M4.5 & 14.30 \nl 
  G 205-019  &  &   18 &  22 &  43.5 &  37 &  57 &  47  & 5 & 9.70 & 24.5 & 7 & 0 & 100 & M1.0 & 11.65 \nl 
  LHS 3420 &  &   18 &  52 &  33.7 &  45 &  38 &  33  & 5 & 13.16 & 24.1 & 7 & 0 & 100 & M4.5 & 15.07 \nl 
  LHS 3429 &  &   19 &   5 &  17.4 &  45 &   7 &  15  & 5 & 12.03 & 23.4 & 5 & 55 & 45 & M3.5 & 13.88 \nl 
  BD+28 3698\tablenotemark{d}&  &   20 &  18 &  13.5 &  29 &  12 &  14  & 5 & 7.42 & 22.7 & 34 & -1 & -1 & K & 9.2\phantom{0} \nl 
  LHS 3547 &  &   20 &  24 &  50.7 &  74 &  12 &  24  & 5 & 14.12 & 17.8 & 5 & 0 & 100 & M5.0 & 15.37 \nl 
  LP 74-35 &  &   20 &  54 &  54.7 &  67 &  35 &   9  & 5 & 10.12 & 24.5 & 7 & 0 & 100 & M1.5 & 12.07 \nl 
  LTT 16250 &  &   21 &  18 &  33.7 &  20 &  57 &   3  & 5 & 10.01 & 25.0 & 42 & -1 & -1 & K & 12.\phantom{0}\phantom{0} \nl 
  LHS 3705\tablenotemark{e}&  &   21 &  44 &  53.9 &  44 &  17 &   8  & 5 & 9.74 & 30.3 & 6 & 14 & 86 & M1.5 & 11.40 \nl 
  G 241-021 &  &   22 &  47 &  47.1 &  66 &  23 &  37  & 5 & 11.89 & 22.1 & 7 & 0 & 100 & M3.5 & 13.61 \nl 
  LHS 3879 &  &   22 &  57 &  41.0 &  37 &  19 &  23  & 5 & 11.45 & 19.3 & 6 & 0 & 100 & M3.0 & 12.88 \nl 
  LHS 3898 &  &   23 &   7 &  46.9 & -27 &  54 &  23  & 5 & 10.38 & 24.8 & 6 & 15 & 85 & M2.5 & 12.35 \nl 
\enddata
\tablenotetext{a}{The source of positional data for all stars is the
Digitized Sky Survey and is denoted by 5.  Sources 1-4 are used in PMSU1.}
\tablenotetext{b}{Based on Fleming {\it et al.} (1988) 
observed $V_{system} = 14.20$ and assumed $\Delta V = 0.42$.}   
\tablenotetext{c}{This is the 809th star in Stephenson (1986).  Photometry is
from Weis (1991).}
\tablenotetext{d}{Jahrei\ss~ notes that this object is presumably a dK binary with 
$\Delta m = 0.8$, see Heintz (1993).  The distance was estimated by Jahrei\ss.}
\tablenotetext{e}{This is also Cou 2234.  Estimated distance assumes
the system is a binary with $\Delta m = 0.0$ as determined 
by Couteau (1985).}  
\end{deluxetable}

\begin{deluxetable}{crrrrrrrrrr}
\tablewidth{0pt}
\tablenum{2}
\label{nearby-bands}
\tablecaption{Candidate Nearby Stars: Bandstrengths}
\tablehead{
\colhead{Star} & 
\colhead{TiO1} & 
\colhead{TiO2} & 
\colhead{TiO3} & 
\colhead{TiO4} & 
\colhead{TiO5} & 
\colhead{CaH1} & 
\colhead{CaH2} & 
\colhead{CaH3} & 
\colhead{CaOH} & 
\colhead{H$\alpha$}
}
\startdata
  LHS 1037 & 0.785 & 0.648 & 0.734 & 0.635 & 0.451 & 0.820 & 0.447 & 0.723 & 0.463 & \nodata \nl 
  LHS 1062 & 0.813 & 0.667 & 0.707 & 0.636 & 0.464 & 0.838 & 0.442 & 0.735 & 0.468 & \nodata \nl 
  LHS 1104 & 0.812 & 0.730 & 0.766 & 0.723 & 0.547 & 0.815 & 0.502 & 0.752 & 0.522 & \nodata \nl 
  LHS 1105 & 0.780 & 0.577 & 0.643 & 0.672 & 0.379 & 0.780 & 0.374 & 0.655 & 0.385 & \nodata \nl 
  G 217-060 & 0.765 & 0.610 & 0.693 & 0.628 & 0.429 & 0.753 & 0.394 & 0.650 & 0.374 & \nodata \nl 
  LHS 1169 & 0.792 & 0.694 & 0.763 & 0.721 & 0.489 & 0.799 & 0.454 & 0.712 & 0.473 & \nodata \nl 
  LHS 1182 & 0.758 & 0.636 & 0.688 & 0.712 & 0.425 & 0.795 & 0.404 & 0.683 & 0.398 & \nodata \nl 
  G 173-018 & 0.854 & 0.735 & 0.809 & 0.776 & 0.589 & 0.808 & 0.516 & 0.742 & 0.547 & 2.92 \nl 
  LHS 1311 & 0.693 & 0.504 & 0.565 & 0.514 & 0.312 & 0.866 & 0.307 & 0.640 & 0.251 & \nodata \nl 
   G 035-027 & 0.753 & 0.511 & 0.639 & 0.570 & 0.371 & 0.775 & 0.305 & 0.608 & 0.299 & 6.12 \nl 
  LP 469-11 & 0.770 & 0.483 & 0.633 & 0.576 & 0.348 & 0.778 & 0.375 & 0.619 & 0.411 & 9.38 \nl 
  LP 709-43   & 0.759 & 0.617 & 0.670 & 0.608 & 0.425 & 0.884 & 0.383 & 0.692 & 0.424 & \nodata \nl 
  LHS 1349 & 0.757 & 1.226 & 1.270 & 1.333 & 1.717 & 0.115 & 0.829 & 0.439 & \nodata & \nodata \nl 
  LHS 1412 & 0.795 & 0.647 & 0.707 & 0.676 & 0.498 & 0.804 & 0.460 & 0.715 & 0.483 & 5.61 \nl 
  G 134-035 & 0.900 & 0.811 & 0.912 & 0.866 & 0.656 & 0.868 & 0.602 & 0.821 & 0.610 & \nodata \nl 
  LHS 1553 & 0.795 & 0.683 & 0.649 & 0.668 & 0.464 & 0.776 & 0.432 & 0.689 & 0.419 & \nodata \nl 
  G 038-024 & 0.832 & 0.774 & 0.806 & 0.741 & 0.645 & 0.753 & 0.410 & 0.702 & 0.404 & \nodata \nl 
  LHS 1631 & 0.749 & 0.640 & 0.678 & 0.650 & 0.454 & 0.781 & 0.424 & 0.680 & 0.419 & \nodata \nl 
  LHS 1659 & 0.793 & 0.685 & 0.737 & 0.701 & 0.487 & 0.855 & 0.471 & 0.752 & 0.486 & \nodata \nl 
  LHS 1675 & 0.783 & 0.638 & 0.695 & 0.619 & 0.427 & 0.803 & 0.455 & 0.732 & 0.450 & \nodata \nl 
  LHS 1679 & 0.710 & 0.470 & 0.595 & 0.529 & 0.307 & 0.761 & 0.319 & 0.604 & 0.321 & 3.39 \nl 
  LHS 1808 & 0.797 & 0.529 & 0.643 & 0.648 & 0.355 & 0.848 & 0.384 & 0.691 & 0.378 & \nodata \nl 
  LHS 1817 & 0.818 & 0.503 & 0.649 & 0.628 & 0.366 & 0.875 & 0.406 & 0.678 & 0.488 & 6.81 \nl 
  LTT 2494 & 0.793 & 0.672 & 0.712 & 0.627 & 0.461 & 0.790 & 0.386 & 0.692 & 0.426 & \nodata \nl 
  LTT 2495 & 0.890 & 0.825 & 0.845 & 0.823 & 0.699 & 0.813 & 0.533 & 0.784 & 0.629 & \nodata \nl 
  LHS 1853 & 0.820 & 0.601 & 0.672 & 0.617 & 0.382 & 0.866 & 0.413 & 0.712 & 0.415 & \nodata \nl 
  LHS 1962 & 0.789 & 0.607 & 0.696 & 0.645 & 0.411 & 0.848 & 0.425 & 0.709 & 0.424 & \nodata \nl 
  LHS 1964 & 0.757 & 0.564 & 0.670 & 0.652 & 0.377 & 0.855 & 0.396 & 0.697 & 0.422 & \nodata \nl 
  LHS 1973 & 0.778 & 0.637 & 0.722 & 0.664 & 0.457 & 0.802 & 0.426 & 0.701 & 0.440 & \nodata \nl 
  LHS 1993 & 0.738 & 0.496 & 0.614 & 0.579 & 0.346 & 0.756 & 0.338 & 0.608 & 0.361 & 3.82 \nl 
  LHS 2002 & 0.695 & 0.504 & 0.585 & 0.512 & 0.303 & 0.747 & 0.315 & 0.608 & 0.275 & \nodata \nl 
  LHS 2025 & 0.778 & 0.611 & 0.689 & 0.636 & 0.413 & 0.829 & 0.419 & 0.714 & 0.408 & \nodata \nl 
  LTT 12375 & 0.826 & 0.766 & 0.801 & 0.753 & 0.584 & 0.818 & 0.591 & 0.801 & 0.616 & \nodata \nl 
  G 253-034 & 0.755 & 0.487 & 0.631 & 0.575 & 0.342 & 0.812 & 0.354 & 0.653 & 0.370 & 5.82 \nl 
  G 042-036 & 0.845 & 0.804 & 0.825 & 0.773 & 0.644 & 0.813 & 0.554 & 0.781 & 0.611 & \nodata \nl 
  LHS 2232 & 0.805 & 0.649 & 0.710 & 0.681 & 0.463 & 0.817 & 0.439 & 0.708 & 0.443 & \nodata \nl 
  Ross 895 & 0.813 & 0.764 & 0.801 & 0.728 & 0.591 & 0.837 & 0.528 & 0.773 & 0.565 & \nodata \nl 
  LHS 2395 & 0.683 & 0.386 & 0.528 & 0.568 & 0.242 & 0.798 & 0.274 & 0.576 & 0.262 & \nodata \nl 
  LHS 2403 & 0.513 & 0.469 & 0.611 & 0.617 & 0.348 & 0.746 & 0.337 & 0.583 & 0.434 & 7.39 \nl 
  Wo 9364 & 0.968 & 0.959 & 1.011 & 0.985 & 0.938 & 0.987 & 0.880 & 0.923 & 0.909 & \nodata \nl 
  LP 63-218 & 0.841 & 0.572 & 0.665 & 0.610 & 0.369 & 0.805 & 0.385 & 0.673 & 0.395 & \nodata \nl 
  LHS 2439 & 0.825 & 0.759 & 0.808 & 0.749 & 0.591 & 0.849 & 0.552 & 0.791 & 0.576 & \nodata \nl 
  LHS 2513 & 0.773 & 0.641 & 0.700 & 0.635 & 0.446 & 0.804 & 0.438 & 0.727 & 0.422 & \nodata \nl 
  Steph 2.809 & 0.789 & 0.625 & 0.703 & 0.706 & 0.467 & 0.760 & 0.419 & 0.654 & 0.475 & 5.94 \nl 
  LHS 2546 & 0.771 & 0.534 & 0.665 & 0.691 & 0.410 & 0.830 & 0.366 & 0.631 & 0.447 & 2.76 \nl 
  LP 321-300 & 0.810 & 0.544 & 0.689 & 0.701 & 0.418 & 0.783 & 0.391 & 0.644 & 0.451 & 10.14 \nl 
  LHS 2613 & 0.797 & 0.602 & 0.677 & 0.623 & 0.427 & 0.806 & 0.415 & 0.691 & 0.451 & 4.58 \nl 
  EXO 1259+1238 & 0.797 & 0.699 & 0.761 & 0.715 & 0.546 & 0.795 & 0.475 & 0.713 & 0.528 & 3.61 \nl 
  LHS 2672 & 0.722 & 0.622 & 0.657 & 0.672 & 0.451 & 0.846 & 0.471 & 0.729 & 0.502 & \nodata \nl 
  LHS 2686 & 0.754 & 0.511 & 0.613 & 0.591 & 0.327 & 0.781 & 0.344 & 0.635 & 0.343 & 2.86 \nl 
  G 014-046 & 1.105 & 1.071 & 1.115 & 0.987 & 0.976 & 1.014 & 0.971 & 1.005 & 0.970 & \nodata \nl 
  LP 271-35 & 0.769 & 0.569 & 0.671 & 0.661 & 0.381 & 0.820 & 0.403 & 0.689 & 0.423 & 1.64 \nl 
  LHS 2935 & 0.793 & 0.415 & 0.619 & 0.619 & 0.299 & 0.818 & 0.306 & 0.609 & 0.331 & 9.7 \nl 
  LHS 3009 & 0.794 & 0.626 & 0.703 & 0.665 & 0.431 & 0.846 & 0.438 & 0.729 & 0.467 & \nodata \nl 
  LHS 3031 & 0.803 & 0.540 & 0.623 & 0.604 & 0.364 & 0.732 & 0.348 & 0.613 & 0.411 & 3.3 \nl 
  LHS 3044 & 0.750 & 0.643 & 0.706 & 0.650 & 0.435 & 0.709 & 0.385 & 0.639 & 0.391 & \nodata \nl 
  GJ 2120 & 0.946 & 0.991 & 0.948 & 1.015 & 0.994 & 1.022 & 0.934 & 0.946 & 0.916 & \nodata \nl 
  G 240-044 & 0.784 & 0.711 & 0.749 & 0.688 & 0.507 & 0.793 & 0.456 & 0.718 & 0.475 & \nodata \nl 
  G 240-045 & 0.739 & 0.608 & 0.670 & 0.625 & 0.394 & 0.757 & 0.378 & 0.655 & 0.378 & \nodata \nl 
  LHS 3318 & 0.739 & 0.541 & 0.638 & 0.556 & 0.345 & 0.809 & 0.365 & 0.660 & 0.371 & 1.39 \nl 
  G 205-019  & 0.849 & 0.817 & 0.861 & 0.818 & 0.674 & 0.855 & 0.610 & 0.818 & 0.658 & \nodata \nl 
  LHS 3420 & 0.774 & 0.523 & 0.624 & 0.594 & 0.359 & 0.754 & 0.349 & 0.619 & 0.360 & \nodata \nl 
  LHS 3429 & 0.772 & 0.656 & 0.712 & 0.658 & 0.434 & 0.778 & 0.411 & 0.678 & 0.424 & \nodata \nl 
  BD+26 3698 & 0.984 & 0.985 & 1.000 & 0.991 & 0.980 & 1.015 & 0.952 & 0.968 & 0.945 & \nodata \nl 
  LHS 3547 & 0.709 & 0.475 & 0.594 & 0.556 & 0.311 & 0.799 & 0.327 & 0.631 & 0.319 & 2.72 \nl 
  LP 74-35 & 0.833 & 0.774 & 0.812 & 0.751 & 0.613 & 0.838 & 0.553 & 0.786 & 0.586 & \nodata \nl 
  LTT 16250 & 0.985 & 0.977 & 1.002 & 0.978 & 0.971 & 0.982 & 0.901 & 0.951 & 0.924 & \nodata \nl 
  LHS 3705 & 0.829 & 0.789 & 0.832 & 0.789 & 0.623 & 0.844 & 0.567 & 0.794 & 0.603 & \nodata \nl 
  G 241-021 & 0.803 & 0.626 & 0.700 & 0.644 & 0.439 & 0.835 & 0.436 & 0.724 & 0.460 & \nodata \nl 
  LHS 3879 & 0.777 & 0.662 & 0.712 & 0.655 & 0.473 & 0.820 & 0.448 & 0.727 & 0.477 & \nodata \nl 
  LHS 3898 & 0.804 & 0.711 & 0.764 & 0.699 & 0.541 & 0.857 & 0.502 & 0.763 & 0.528 & \nodata \nl 
\enddata
\end{deluxetable}

\begin{deluxetable}{crrrrrrr}
\tablewidth{0pt}
\tablenum{3}
\label{nearby-uvw}
\tablecaption{Candidate Nearby Stars: Space Motions}
\tablehead{
\colhead{Star } & 
\colhead{$\mu_\alpha$} & 
\colhead{$\mu_\delta$} & 
\colhead{$V_{rad}$} &
\colhead{U} & 
\colhead{V} & 
\colhead{W} &
\colhead{$M_V$}  
}
\startdata
  LHS 1037 & -0.5755 & -0.3896 & 1.0 & 54. & 12. & -22. & 11.72 \nl 
  LHS 1062 & 0.6123 & -0.3093 & -62.6 & -12. & -85. & -4. & 11.56 \nl 
  LHS 1104 & 0.7723 & -0.1613 & -25.6 & -54. & -66. & -17. & 10.80 \nl 
  LHS 1105 & 0.7723 & -0.1613 & 26.3 & -87. & -26. & -29. & 13.14 \nl 
  G 217-060 & -0.4107 & -0.2286 & -47.4 & 59. & -22. & -13. & 12.73 \nl 
  LHS 1169 & 0.6397 & -0.0548 & -13.6 & -27. & -35. & 2. & 11.26 \nl 
  LHS 1182 & 0.5219 & -0.0338 & -85.2 & 0. & -101. & -14. & 12.08 \nl 
  G 173-018 & 0.4100 & 0.2358 & 11.5 & -33. & -6. & 21. & 10.31 \nl 
  LHS 1311 & 0.6324 & -0.2710 & -2.5 & -27. & -53. & 9. & 14.10 \nl 
  G 035-027 & 0.2471 & 0.0303 & -11.7 & -13. & -19. & 18. & 12.95 \nl 
  LP 469-118 & 0.2471 & 0.0303 &  7.8 & -25. & -12. & 5. & 13.37 \nl 
  LP 709-43   & -0.3479 & -0.2527 & -17.3 & 33. & 1. & 3. & 11.29 \nl 
  LHS 1349 & 0.0363 & -0.9903 & 0.0 & -6. & -35. & -80. & 14.37 \nl 
  LHS 1412 & 0.5562 & -0.0732 & 39.7 & -54. & -52. & -14. & 11.16 \nl 
  G 134-035 & 0.2560 & 0.0045 & 23.6 & -31. & 1. & 0. & 12.12 \nl 
  LHS 1553 & 0.3710 & -0.4905 & -27.1 & -33. & -65. & -27. & 11.56 \nl 
  G 038-024 & 0.3290 & -0.1534 & -19.9 & 4. & -44. & 19. & 10.71 \nl 
  LHS 1631 & 0.6737 & -0.5877 & -29.1 & -43. & -85. & -4. & 11.69 \nl 
  LHS 1659 & 0.0496 & -0.5247 & -58.8 & 48. & -60. & -26. & 10.97 \nl 
  LHS 1675 & -0.1884 & -0.4857 & -44.7 & 55. & -1. & 4. & 12.96 \nl 
  LHS 1679 & 0.2049 & -0.5852 & 8.2 & -48. & -37. & -22. & 14.21 \nl 
  LHS 1808 & 0.3432 & -0.4848 & 45.6 & -60. & -21. & 26. & 13.24 \nl 
  LHS 1817 & 0.3224 & -0.7746 & -19.4 & -8. & -56. & -7. & 13.04 \nl 
  LTT 2494 & 0.2208 & -0.1275 & 48.3 & -31. & -43. & 3. & 11.69 \nl 
  LTT 2495 & -0.0071 & 0.2039 & 40.4 & -45. & -6. & 2. & 10.74 \nl 
  LHS 1853 & 0.3049 & -0.5479 & 62.5 & -73. & -20. & 34. & 12.76 \nl 
  LHS 1962 & -0.2874 & -0.4654 & -23.4 & -38. & -39. & -32. & 12.29 \nl 
  LHS 1964 & 0.1681 & -0.7166 & -48.4 & 35. & -48. & -21. & 12.84 \nl 
  LHS 1973 & 0.4841 & -0.2206 & -96.4 & 95. & -67. & -1. & 11.65 \nl 
  LHS 1993 & -0.3181 & -0.4101 & -2.1 & -36. & -24. & -18. & 13.41 \nl 
  LHS 2002 & -0.4800 & 0.1504 & 68.7 & -82. & -8. & 1. & 14.30 \nl 
  LHS 2025 & 0.5973 & 0.3123 & -106.0 & 99. & -44. & -33. & 12.42 \nl 
  LTT 12375 & -0.0339 & -0.2143 & -5.9 & 10. & -20. & -15. & 10.35 \nl 
  G 253-034 & -0.1259 & -0.3458 & -12.3 & -15. & -37. & 6. & 13.48 \nl 
  G 042-036 & -0.3036 & -0.0159 & 5.5 & -33. & -10. & -19. & 9.90 \nl 
  LHS 2232 & -0.3836 & -0.5073 & -17.3 & -6. & -33. & -16. & 11.57 \nl 
  Ross 895 & 0.1712 & 0.0727 & -66.3 & 32. & 53. & -36. & 9.77 \nl 
  LHS 2395 & 0.3023 & -0.6225 & -49.0 & 40. & -30. & -31. & 15.81 \nl 
  LHS 2403 & -0.5448 & -0.2290 & 32.7 & -55. & -39. & 17. & 13.37 \nl 
  Wo 9364 & -0.1712 & -0.0361 & 24.0 & -25. & -10. & 17. & 7.81 \nl 
  LP 63-218 & 0.0283 & -0.2010 & 21.6 & -1. & -13. & 33. & 12.98 \nl 
  LHS 2439 & -0.5448 & 0.0152 & -34.5 & -27. & -40. & -36. & 10.29 \nl 
  LHS 2513 & -0.6518 & -0.3628 & 32.0 & -66. & -35. & 37. & 11.79 \nl 
  Steph 2.809 & -0.1113 & 0.0450 & -18.7 & -3. & -7. & -19. & 11.52 \nl 
  LHS 2546 & -0.3675 & -0.3637 & 4.1 & -28. & -70. & 22. & 12.31 \nl 
  LP 321-300 & -0.1294 & -0.0718 & -5.0 & -9. & -16. & -6. & 12.18 \nl 
  LHS 2613 & -0.5495 & 0.0404 & -27.0 & -21. & -19. & -28. & 12.15 \nl 
  EXO 1259+1238 & -0.1876 & -0.0875 & -1.6 & -14. & -21. & -3. & 10.68 \nl 
  LHS 2672 & -0.5488 & -0.1847 & 6.4 & -30. & -37. & 12. & 11.72 \nl 
  LHS 2686 & -0.5974 & -0.5689 & -15.8 & -9. & -56. & 1. & 13.79 \nl 
  G 014-046 & 0.1353 & -0.2655 & 105.5 & 63. & -47. & 77. & 9.88 \nl 
  LP 271-35 & 0.0178 & -0.2032 & -19.6 & 14. & -21. & -18. & 12.77 \nl 
  LHS 2935 & -0.5016 & 0.0210 & 0.9 & -29. & -27. & 18. & 14.39 \nl 
  LHS 3009 & -0.6374 & 0.3150 & -60.6 & -56. & -68. & -30. & 12.00 \nl 
  LHS 3031 & 0.3881 & -0.3841 & -22.0 & 64. & -4. & -33. & 13.07 \nl 
  LHS 3044 & -0.3994 & 0.3738 & 1.0 & -64. & -18. & 6. & 11.94 \nl 
  GJ 2120 & -0.1934 & -0.3095 & 79.6 & 75. & -24. & 32. & 9.64 \nl 
  G240-044 & -0.1297 & 0.1923 & -52.3 & -23. & -51. & -21. & 11.06 \nl 
  G240-045 & -0.1297 & 0.1923 & -54.8 & -25. & -54. & -21. & 12.56 \nl 
  LHS 3318 & 0.5747 & 0.2816 & -7.3 & -18. & 11. & -41. & 13.43 \nl 
  G 205-019  & 0.0262 & -0.3741 & -57.8 & 17. & -60. & -37. & 9.70 \nl 
  LHS 3420 & 0.2156 & 0.4666 & -38.4 & -65. & -19. & -17. & 13.16 \nl 
  LHS 3429 & 0.0170 & -0.6088 & -57.6 & 48. & -61. & -43. & 12.03 \nl 
  BD+28 3698 & -0.0020 & -0.0900 & 40.4 & 22. & 34. & -8. & 7.42 \nl 
  LHS 3547 & 0.2369 & 0.6171 & -18.4 & -46. & -36. & 2. & 14.12 \nl 
  LP 74-35 & 0.1758 & 0.1529 & -18.8 & -22. & -23. & -9. & 10.12 \nl 
  LTT 16250 & 0.0892 & -0.2325 & 44.4 & 24. & 27. & -39. & 10.01 \nl 
  LHS 3705 & -0.1347 & -0.6390 & -40.9 & 75. & -46. & -52. & 9.74 \nl 
  G 241-021 & 0.2670 & 0.3758 & -20.6 & -32. & -37. & 20. & 11.89 \nl 
  LHS 3879 & -0.5405 & -0.3483 & -75.8 & 69. & -63. & 20. & 11.45 \nl 
  LHS 3898 & 0.7667 & -0.0982 & -0.3 & -74. & -40. & -35. & 10.38 \nl 
\enddata
\end{deluxetable}

\begin{deluxetable}{llrrrrrcr}
\tablewidth{0pt}
\tablenum{4}
\label{lhs-spec}
\tablecaption{Faint LHS Stars}
\tablehead{
\colhead{Star} & 
\colhead{$m_r$} & 
\colhead{TiO5} & 
\colhead{CaH1} &
\colhead{CaH2} &
\colhead{CaH3} & 
\colhead{H$\alpha$}&
\colhead{Sp. Type} & 
\colhead{$V_{rad}$} 
}
\startdata
 LHS 1024 & 18.0 & 0.698 & 0.533 & 0.377 & 0.592 & \nodata &   sdM4.0  & -60  \nl 
 LHS 1035 & 17.7 & 0.221 & 0.572 & 0.246 & 0.506 &\nodata &   sdM6.0  & -63  \nl 
 LHS 1055 & 17.2 & 0.311 & 0.935 & 0.343 & 0.652 &\nodata &   M4.5 V  & 19  \nl 
 LHS 1109 & 17.7 & 0.320 & 0.763 & 0.326 & 0.603 &\nodata &   M5.0 V  & -1  \nl 
 LHS 1135 & 18.7 & 0.194 & 0.618 & 0.214 & 0.471 &\nodata &   sdM6.5  & -27  \nl 
 LHS 1146 & 18.2 & 0.191 & 0.853 & 0.247 & 0.591 & 5.7 &   M6.0 V  & 58  \nl 
 LHS 1157 & 19.0 & 0.187 & 0.806 & 0.272 & 0.625 &\nodata &   M5.5 V  & -120  \nl 
 LHS 1294\tablenotemark{a} & 18.0 & 0.655 & 0.966 & 0.567 & 0.834 &\nodata &   M1.5 V  & 24  \nl 
 LHS 1506 & 18.0 & 0.253 & 0.790 & 0.274 & 0.580 &\nodata &   M5.5 V  & \nodata  \nl 
 LHS 1531 & 17.6 & 0.182 & 0.699 & 0.206 & 0.486 &\nodata&   M6.0 V  & \nodata  \nl 
 LHS 1657\tablenotemark{a}& 19.0 & 0.559 & 0.776 & 0.473 & 0.740 & 2.1 &   M2.5 V  & 35  \nl 
 LHS 1669 & 17.7 & 0.464 & 0.716 & 0.356 & 0.630 &\nodata &   sdM4.5  & \nodata  \nl 
 LHS 1826 & 18.4 & 1.079 & 0.333 & 0.260 & 0.363 &\nodata &   esdM6.0  & 177  \nl 
 LHS 1937 & 18.1 & 0.268 & 0.862 & 0.255 & 0.583 & 3.7 &   M7.0 V  & \nodata  \nl 
 LHS 1981 & 17.8 & 0.707 & 0.617 & 0.420 & 0.637 &\nodata &   sdM3.5  & 78  \nl 
 LHS 1986\tablenotemark{a}& 17.7 & 0.989 & 1.040 & 0.952 & 1.033 &\nodata &   K  & \nodata  \nl 
 LHS 2064 & 17.6 & 0.224 & 0.724 & 0.259 & 0.547 &\nodata &   M6.0 V  & 36  \nl 
 LHS 2089 & 18.0 & 0.898 & 0.724 & 0.560 & 0.713 &\nodata &   esdM1.5  & -2  \nl 
 LHS 2099 & 16.0 & 0.956 & 0.698 & 0.517 & 0.711 &\nodata &   esdM2.0  & \nodata  \nl 
 LHS 2100 & 19.0 & 0.894 & 0.682 & 0.499 & 0.681 &\nodata &   esdM2.5  & \nodata  \nl 
 LHS 2139 & 18.5 & 0.894 & 1.136 & 0.836 & 0.936 &\nodata &   DC?  & \nodata  \nl 
 LHS 2140 & 14.8 & 0.893 & 0.770 & 0.663 & 0.820 &\nodata &   sdM0.5  & -27  \nl 
 LHS 2167 & 18.5 & 0.272 & 0.814 & 0.285 & 0.586 &\nodata &   M5.5 V  & \nodata  \nl 
 LHS 2174 & 17.9 & 0.638 & 0.683 & 0.420 & 0.630 &\nodata &   sdM3.5  & -101  \nl 
 LHS 2187 & 18.6 & 0.689 & 0.574 & 0.411 & 0.605 &\nodata &   sdM3.5  & 172  \nl 
 LHS 2195 & 19.2 & 0.309 & 0.920 & 0.336 & 0.620 & 2.7 &   M8.0 V  & 50  \nl 
 LHS 2236 & 18.0 & 0.311 & 0.712 & 0.275 & 0.542 &\nodata &   sdM5.5  & \nodata  \nl 
 LHS 2300 & 17.9 & 0.663 & 0.647 & 0.363 & 0.573 &\nodata &   sdM4.5  & \nodata  \nl 
 LHS 2347 & 17.7 & 0.302 & 0.849 & 0.293 & 0.625 &\nodata &   M5.5 V  & \nodata  \nl 
 LHS 2419 & 17.8 & 0.314 & 0.701 & 0.284 & 0.573 &\nodata &   sdM5.5  & \nodata  \nl 
 LHS 3681 & 17.7 & 0.891 & 0.596 & 0.452 & 0.672 &\nodata &   esdM3.0  & 42  \nl 
 LHS 3762 & 17.9 & 0.251 & 0.852 & 0.299 & 0.656 &\nodata &   M5.0 V  & -46  \nl 
 LHS 3868 & 17.7 & 0.970 & 0.835 & 0.676 & 0.833 &\nodata &   esdM0.5  & -49  \nl 
 LHS 3958 & 18.0 & 0.579 & 0.585 & 0.356 & 0.570 &\nodata &   sdM4.5  & -231  \nl 
 LHS 5071 & 17.7 & 0.453 & 0.842 & 0.337 & 0.626 &\nodata &   sdM4.5  & 56  \nl 
 LHS 5081 & 11.7 & 0.657 & 0.845 & 0.592 & 0.808 &\nodata &   M1.5 V  & 37  \nl 
 LHS 5356 & 17.6 & 0.264 & 0.794 & 0.296 & 0.603 &\nodata &   M5.5 V  & -18  \nl 
 LHS 5359 & 14.7 & 0.586 & 0.778 & 0.520 & 0.766 &\nodata &   M2.0 V  & -131  \nl 
 LHS 5360 & 18.0 & 0.369 & 0.693 & 0.352 & 0.623 &\nodata &   M4.5 V  & -126  \nl 
 LHS 5387 & 17.6 & 0.418 & 0.733 & 0.380 & 0.651 &\nodata &   M4.0 V  & -210  \nl 
\enddata
\tablenotetext{a}{May be misidentified}
\end{deluxetable}

\clearpage

%--------------------------BIBLIOGRAPHY---------------------------

\clearpage

\end{document}